\documentclass{paper}

\usepackage{graphicx} 

\title{On ``Nonlinear eigenvalue problems'' }
\author{Oliver S. Kerr\\Department of Mathematics, City University London,\\Northampton Square, London, EC1V 0HB, U.K.}

\begin{document}
\maketitle

\abstract{The asymptotic behaviour of solutions to $y'(x)=\cos[\pi x y(x)]$ was investigated by Bender, Fring and Komijani \cite{BenderEtAl:2014}. They found, for example, a relation between the initial value $y(0)=a$ and the number of maxima that the solution exhibited. We present an alternative derivation of the asymptotic results that looks at the solutions in the regions $x<y$ and $x>y$, and confirms the behaviour found previously for larger values of $a$. This method uses the small amplitude and high frequency of the oscillatory behaviour in the region $x<y$.}

\bigskip\bigskip

\newpage

\section{Introduction}

In a recent paper by Bender, Fring and Komijani \cite{BenderEtAl:2014} a detailed asymptotic analysis of the nonlinear initial-value problem 
\begin{equation}
y'(x)=\cos[\pi x y(x)],\quad y(0)=a
\label{Prob}
\end{equation} 
was presented which focused on the solutions for $x\ge 0$. They showed that for $a>0$ solutions could be split into classes depending on the initial conditions such that solution with $a_{n-1}<a<a_n$ displayed an oscillatory region with $n$ maxima before decaying monotonically to zero. They then found the result that as $n\to\infty$, $a_n\sim 2^{5/6}\sqrt{n}$. The purpose of this paper is to present an alternative derivation of this result that may be simpler to adapt to analogous problems.

\section{Outline}
\begin{figure}[h]
\centerline{\hspace*{0.4in}\raisebox{2.24in}{\makebox[0in][l]{\hspace{0.1in}$y$}}\includegraphics[width=4.6in]{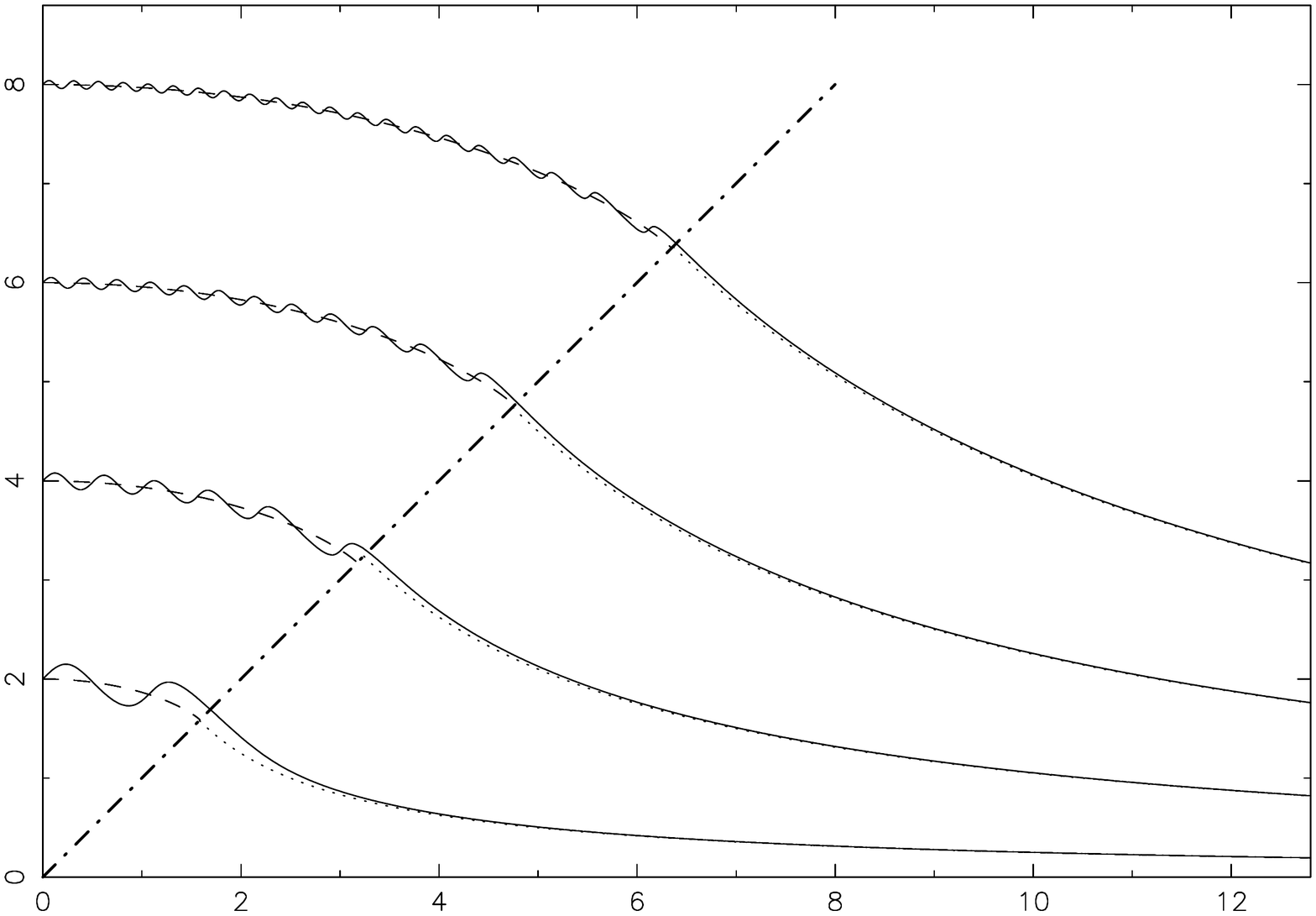}\raisebox{0.12in}{\makebox[0in][l]{\hspace{-1.75in}$x$}}}
\caption{Plots of solutions to (\ref{Prob}) with $y(0)=2,\,4,\,6,\,8$. The dotted lines in $x>y$ show the curves $xy=C$ to which these converge asymptotically as $x\to\infty$. The dashed lines in $x<y$ give the estimate of the mean path of the oscillatory part of these curves.}
\label{ExamplePlot}
\end{figure}

The typical behaviour of solutions to (\ref{Prob}) is shown by the solid lines in figure~\ref{ExamplePlot}. There is an initial oscillatory phase where the frequency increases and the amplitude decreases as the initial value, $y(0)$, increases. These oscillatory solutions drift downwards until they undergo a transition to monotonic decay towards the horizontal axis.

\begin{figure}[h]
\centerline{\hspace*{-0.3in}\raisebox{2.24in}{\makebox[0in][l]{\hspace{0.16in}$y$}}\includegraphics[width=4.6in]{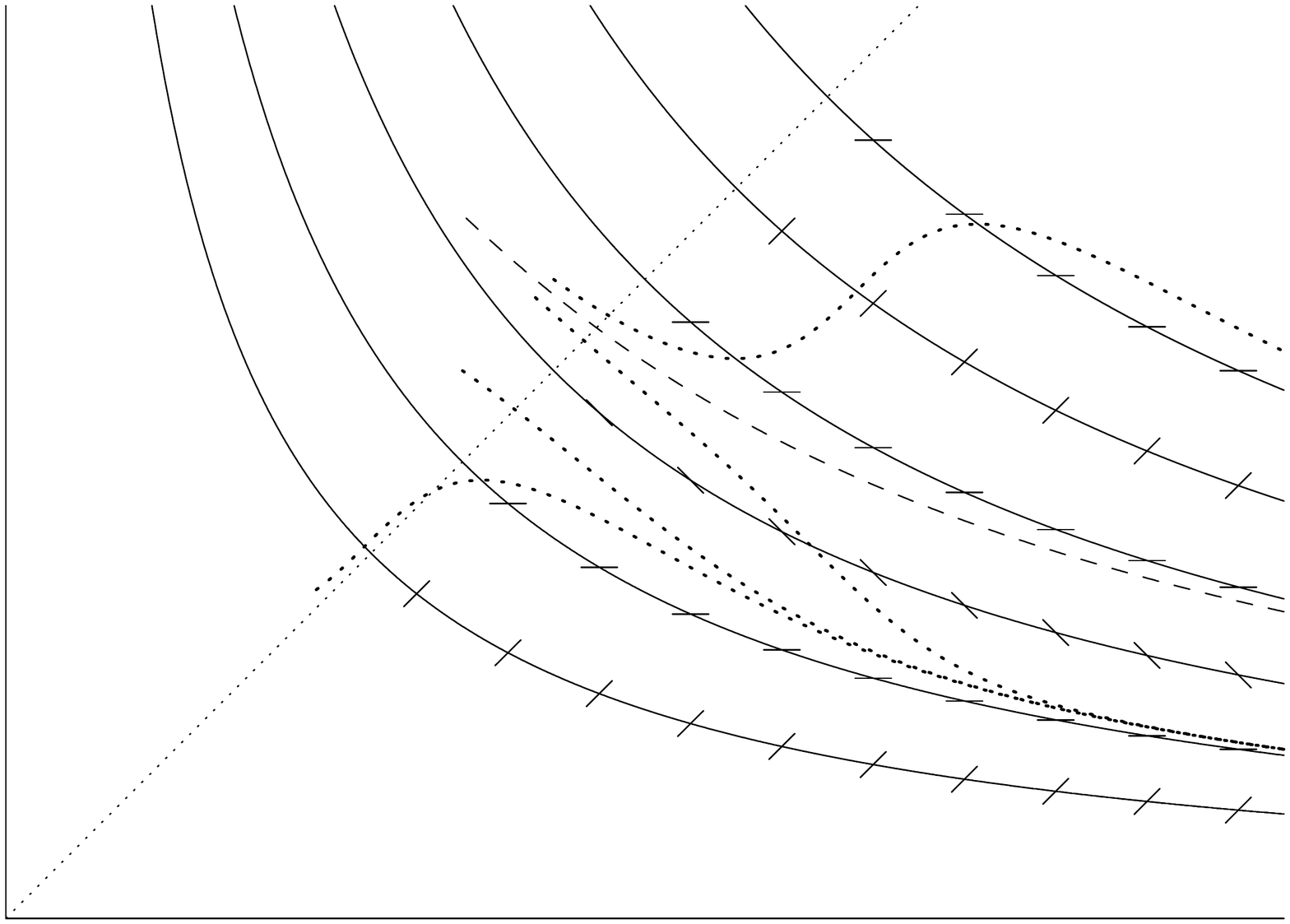}\raisebox{0.12in}{\makebox[0in][l]{\hspace{-1.75in}$x$}}\hspace{-0.6in}\raisebox{2.6in}{\makebox[0in][l]{\hspace*{-1.2in}$x=y$}}\raisebox{0.58in}{\makebox[0in][l]{$xy=2n$}}\raisebox{0.75in}{\makebox[0in][l]{$xy=2n+\frac{1}{2}$}}\raisebox{0.94in}{\makebox[0in][l]{$xy=2n+1$}}\raisebox{1.16in}{\makebox[0in][l]{$xy=2n+\frac{3}{2}$}}\raisebox{1.4in}{\makebox[0.0in][l]{$xy=2n+2$}}\raisebox{1.7in}{\makebox[0.8in][l]{$xy=2n+\frac{5}{2}$}}}
\caption{Schematic plot for solutions in the region $x>y$ and the influences of the lines of form $xy=C$. The dotted lines show the trajectories of various solutions, while the dashed line shows the path of the separatrix dividing solutions that converge to $xy=2n+1/2$ from those that converge to $xy=2n+5/2$.}
\label{xyPlot}
\end{figure}

Some of the basic behaviour of the solutions of (\ref{Prob}) can be understood by considering the lines in the $x$--$y$ plane where $xy$ is constant. The situation is shown schematically in figure~\ref{xyPlot}.
 If we consider lines where $xy=2n$ then solutions will have gradient $1$ where they intersect these lines, similarly when $xy=2n+1$ they will intersect with gradient $-1$, and when $xy=2n\pm1/2$ they will intersect with gradient $0$. The gradients of the solutions will have gradients with magnitude at most $1$, while the lines $xy=c$ for constants $c$ have gradients greater than 1 in magnitude for $x<y$, and less than $1$ for $x>y$. In the region $x<y$ solutions must cross the lines $xy=c$ from left to right, with a maximum each time it crosses a line $xy=2n+\frac{1}{2}$, $n=1,\,2,\,3,\,\ldots$. 
In the region $x>y$ this restriction no longer holds. This results in the solutions having intrinsically different behaviour above and below the line $x=y$.

\section{Solutions in the region $x>y$}

The behaviour of solutions in this area is addressed by problem 4.13 in the book by Bender and Orszag \cite{BenderOrszag}. Any solution that enters a region $2n+\frac{1}{2}\le xy\le 2n+1$ is trapped in this region as $x$ increases as the gradient of a solution on the lower boundary is $0$, and on the upper boundary is $-1$. Indeed, in such a region any solution that is initially above a line $xy=2n+\frac{1}{2}+\epsilon$ will have a negative gradient of magnitude greater than $\sin\pi\epsilon$ and so must eventually pass below $xy=2n+\frac{1}{2}+\epsilon$, whose gradient tends to zero as $x\to\infty$. Thus all solutions in this region asymptote to the lines $xy=2n+\frac{1}{2}$. 

All solutions in the region $2n-\frac{1}{2} < xy < 2n+\frac{1}{2}$ will have positive gradients and so will pass into the region $2n+\frac{1}{2}\le xy\le 2n+1$ from below, and will have one maximum in the region $x>y$.

As Bender {\it et al.} showed, there is one solution in the region $2n+1 < xy < 2n+\frac{3}{2}$ that stays in this region. Solutions initially below this curve will pass into the region the curve gradients of any solution $2n+\frac{1}{2}\le xy\le 2n+1$ and remain there, while those above it will end up in the region $2n+\frac{5}{2}\le xy\le 2n+3$. This curve is indicated by the dashed line in figure~\ref{xyPlot}. By a similar argument to that given previously it can be shown that these seperatrices tend towards their asymptotes $xy=2n+3/2$ from below. We will denote the point where the separatrix crosses the line as $x=y=b_n$, and hence $(2n+1)^{1/2}<b_n<(2n+3/2)^{1/2}$.

Clearly, any solution that ends up just above the curve $xy=2n+\frac{1}{2}$ will have crossed $n$ lines given by $xy=\frac{1}{2},\,\frac{5}{2},\,\frac{9}{2},\,\ldots$ and so will have $n$ maxima. Hence any solution that crosses the line $x=y$ with $b_{n-1}<x=y<b_n$ will have $n$ maxima.

\section{Solutions in the region $x<y$}

\begin{figure}[h]
\centerline{\hspace*{0.4in}\raisebox{2.24in}{\makebox[0in][l]{\hspace{0.67in}$y$}}\includegraphics[width=4.2in]{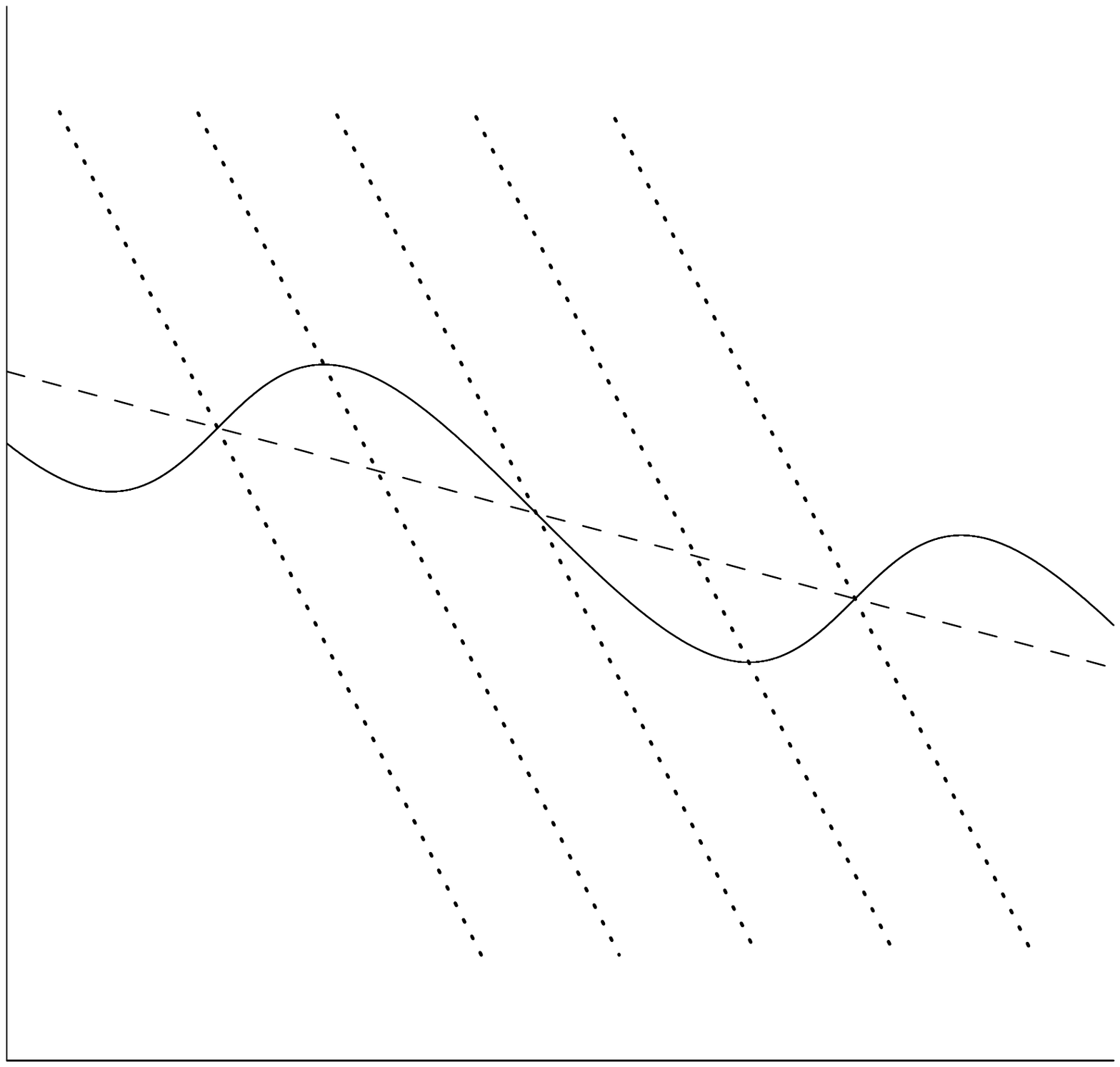}\raisebox{0.12in}{\makebox[0in][l]{\hspace{-1.75in}$x$}}}
\caption{Diagram showing the local approximation to lines of $xy=C$ as parallel lines of the form $x+\alpha y=D$ (dotted lines for $D=2n,\,2n+1/2,\,2n+1,\,2n+3/2,\,2n+2$). The solution (solid line) drifts downwards on average (dashed line).}
\label{Parallel}
\end{figure}

For large values of $y$ the solution $y(x)$ will tend to oscillate quickly with small amplitude. What we are going to do is exploit this to seek the behaviour of the local average of the solution. A schematic diagram of the behaviour of an enlarged section of the trajectory is shown in figure~\ref{Parallel}.

Locally we can approximate the lines of constant $x$ as parallel lines, as shown in figure~\ref{Parallel}.
 and so we can determine the leading order behaviour by looking at the solutions to
\begin{equation}
y'(x)=\cos[\omega(x+\alpha y)],\quad 0<\alpha<1.
\end{equation}

If we let
\begin{equation}
X=x+\alpha y
\end{equation}
then
\begin{equation}
\frac{{\rm d}y}{{\rm d}X}=\frac{ \frac{{\rm d}y}{{\rm d}x}}{1+\alpha\frac{{\rm d}y}{{\rm d}x}}=\frac{\cos\omega X}{1+\alpha\cos\omega X}
\end{equation}
We are interested in the average change in $y$ over a cycle, say $\Delta y$, which corresponds to $X$ changing by $2\pi/\omega$. This is given by
\begin{equation}
\Delta y=\int_0^{2\pi/\omega}\frac{\cos\omega X}{1+\alpha\cos\omega X}\,{\rm d}X.
\end{equation}
This can be evaluated by, say, converting the integral into a complex contour integral about the unit circle centred on the origin, $C$,  by setting $z=e^{i\omega X}$ to give
\begin{equation}
\Delta y=\frac{1}{i\omega}\oint_C\frac{z^2+1}{z(\alpha z^2+2z+\alpha)}\,dz
\end{equation}
Evaluating this using residues gives
\begin{equation}
\Delta y=\frac{2\pi}{\omega}\left(\frac{1}{\alpha}-\frac{1}{\alpha\sqrt{1-\alpha^2}}\right).
\end{equation}
As $X$ changes by $\Delta X =2\pi/\omega$, this gives the average slope of the path to be
\begin{equation}
\frac{\Delta y}{\Delta X}=\frac{1}{\alpha}-\frac{1}{\alpha\sqrt{1-\alpha^2}},
\end{equation}
which is independent of the frequency of the oscillations. At leading order we can use this to give the slope of the mean line 
\begin{equation}
\frac{{\rm d}y}{{\rm d}X}=\frac{1}{\alpha}-\frac{1}{\alpha\sqrt{1-\alpha^2}}.
\end{equation}
where $\alpha$ changes on a scale longer than that of the oscillations.
Converting back to $x$--$y$ coordinates gives
\begin{equation}
\frac{{\rm d}y}{{\rm d}x}=\frac{ \frac{{\rm d}y}{{\rm d}X}}{1-\alpha\frac{{\rm d}y}{{\rm d}X}}=\frac{\sqrt{1-\alpha^2}-1}{\alpha}.
\end{equation}

Up to now we have not specified what $\alpha$ is. It is determined by the hyperbolas $xy=c$. On such curves
\begin{equation}
\frac{{\rm d}y}{{\rm d}x}=-\frac{c}{x^2}=-\frac{y}{x}.
\end{equation}
Since the gradient of the lines $x+\alpha y= c$ is $-1/\alpha$, we find $\alpha=x/y$ and so the equation for the slope of the average curve is given by
\begin{equation}
\frac{{\rm d}y}{{\rm d}x}=\sqrt{\left(\frac{y}{x}\right)^2-1}-\frac{y}{x}.
\end{equation}
This has solutions
\begin{equation}
y^3+3x^2y+\left(y^2-x^2\right)^{3/2}=2y(0)^3.
\end{equation}
The solution curves meet the line $x=y$ when
\begin{equation}
x=y=2^{-1/3}y(0)
\end{equation}
These curves are shown by the dashed lines in figure~\ref{ExamplePlot}, showing good agreement even for these moderate values of $y(0)$.

The $a_n$ of Bender {\it et al.} are the values of $y(0)$ which correspond to the $b_n$, and so to leading order
\begin{equation}
a_n\approx 2^{1/3}b_n=2^{1/3}\left(2^{1/2}n^{1/2}+O(n^{-1/2})\right)=2^{5/6}n^{1/2}+O(n^{-1/2}),
\end{equation}
giving the same result as found by Bender {\it et al.}.

\section{Conclusions}

We have presented here alternative derivation of the results of Bender, Fring and Komijani \cite{BenderEtAl:2014}. Their derivation could be considered to be more mathematically elegant, and for this problem easier to obtain higher order asymptotic behaviour than it would be by the methods presented here. However, the alternative approach adopted here obtains the same leading order behaviour in a way that is possibly more adaptable to analogous problems, and where the leading order asymptotic behaviour is sufficient.

\bibliography{OnNEP}
\bibliographystyle{hplain}

\end{document}